\documentclass[11pt]{article}
\usepackage{sao1}
\usepackage{graphicx}
\def\iue{$IUE$}
\def\units{$10^{-11}\mathrm{erg}\hspace{0.8mm}\mathrm{s}^{-1}\mathrm{cm}^{-2}\mathrm{\AA}^{-1}$}
\begin{document}
%
%
\pagestyle{plain} \setcounter{page}{348}
\title{Ultraviolet variability of the mCP star 56~Arietis
\thanks{Based on $INES$ data from the $IUE$ satellite.}}
\author{N. A. Sokolov}
\institute{Central Astronomical Observatory at Pulkovo,
St. Petersburg 196140, Russia \\ e-mail:~sokolov@gao.spb.ru}
\maketitle
\begin{abstract}
The spectrophotometric variability of the mCP star 56~Ari
in the far-UV spectral region from 1150~\AA\ to 1980~\AA\ is investigated.
This study is based on the archival {\it IUE} data obtained at
different phases of the rotational cycle.
The brightness of 56~Ari is not constant in the investigated wavelengths
over the whole rotational period.
The monochromatic light curves continuously change their shape with wavelength.
The comparison of energy distributions at three phases show that
the first minimum of light curves at phase 0.25 is replaced by the maximum
for $\lambda~>$~1608~\AA, but the second minimum of light curves at
phase 0.65 absent in the spectral region between $\lambda$1938~\AA\
and $\lambda$1980~\AA.
The variable broad features in the far-UV connected with a
non-uniform distribution of silicon over the surface of 56~Ari
influence substantially the light variations in the UV.


\end{abstract}

\section {Introduction}
The mCP~star 56 Arietis (SX Ari, HD~19832 HR~954) belongs to the Si-group.
It shows light, spectrum
and magnetic field variations with a relatively short period of 0.728 days
first established by Provin~(1953).
Jamar~(1978) investigated the ultraviolet spectral variations
of six CP Si stars, including 56~Ari, by means of observations obtained
from the Sky Survey Telescope (S2/68) on the TD1 satellite.
In the case of 56~Ari, the author established that the short wavelength
region is clearly in phase opposition with the remaining part of
the spectrum with a null wavelength region at 1600~\AA.
Sonneborn \& Panek~(1984) and Stepie{\' n} \& Czechowski~(1993) analysed
the spectrophotometric behavior
of this star, using the {\it International Ultraviolet Explorer\/}
({\iue}) data and the published visual spectrophotometric data.
Sonneborn \& Panek confirmed that the spectrophotometric variations
above and below 1600~\AA\ are roughly antiphased.
The authors mentioned that the light curves at the various wavelengths
show remarkable differences in exact phase and shape which indicate that
the mechanism of flux redistribution is complex.
On the other hand,
Stepie{\' n} \& Czechowski pointed out that the variations in the visual
spectral region are in antiphase to the UV variations, but there exists
no null region in the spectrum of 56~Ari.
Instead, the light curve changes continuosly its shape.
Unfortunately the authors are restricted to a qualitative comparison
of light variations in selected spectral bands only with variations
of silicon and helium in visual spectral region.
In the far-UV there are a number of features and
spectral lines that are mainly responsible for variability of the fluxes
in the spectra of CP Si stars.
This may be due to the fact that authors have called the light curves
"monochromatic", although they were determined by averaging the fluxes
over intervals [$\lambda$-10,~$\lambda$+10]~\AA\ for given $\lambda$.
Moreover, the authors investigated only two series of the low-dispersion
spectra obtained in December 1981 (17 SWP spectra) and other in
February 1984 (7 SWP spectra).
It should be that they did not study the variation of
the flux around 1780~\AA.
This spectral feature may be also due to Si~II, but this feature is not
identified yet.

Many improvements were made to the standard processing of {\iue}
data along the years.
The more relevant modifications are: 1) the use of a new noise model,
2) a more accurate representation of the spatial profile of the spectrum,
and 3) a more reliable determination of the background.
These, together with other
modifications, were taken into account in the {\it INES\/} system
developed by the {\it ESA\/} {\iue} Observatory. The $INES$ data from
{\iue} satellite are available from the {\it INES\/} Principal Center
{\rm http://ines.vilspa.esa.es} or from the {\it INES\/} National Hosts
(Wamsteker~2000).

In this paper, the low-dispersion spectra of CP Si star 56~Ari are analysed
using the final {\iue} archive.
Moreover, the variability of selected features in the far-UV can be
established.

\section {Observational data}

\subsection{{\iue} spectra}

Three series observations of 56~Ari obtained with Short Wavelength
Prime (SWP) camera were received from the final {\iue} archive:
\begin{itemize}
\item the first one contains 19 low-dispersion SWP spectra obtained
in December 1981,
\item the second one contains 13 low-dispersion and 5 high-dispersion
SWP spectra obtained in February 1984,
\item the third one contains 8 low-dispersion and 6 high-dispersion
SWP spectra obtained in August and September 1990.
\end{itemize}
In all cases, the large-aperture observational data were held.

In the {\iue} archive, each high-dispersion image has an associated
"rebinned" spectrum, which is obtained by rebinning the "concatenated"
spectrum at the same wavelength step size (1.6764~\AA/pixel) as
low resolution data (Gonz{\'a}lez-Riestra et al.~2000).
This data set represents an important complement to the low resolution
archive, and it is especially useful for time variability studies.
In our study, the "rebinned" spectra from high-dispersion images of 56~Ari
were used.
Finally, we analysed 49 SWP spectra, distributed quite smoothly over
the rotational period.

\section{The period variations of 56~Ari}

The rotational period of 56~Ari has been studied by various
investigators over 48 years (see, e.g., Adelman et al.~(2001)).
Musielok~(1988) found the increase in the rotational period
of 4~s per 100 years from analysis of the $UBV$ and $uvby\beta$ photometric
data. Adelman \& Fried~(1993) and {\v Z}i{\v z}novsk{\' y} et al.~(2000)
noted slight variations of the amplitude and shapes even within
the same photometric band.
Recently, Adelman et al.~(2001) studied all possible variations
for this star from 1952 to 2000. They confirmed the increase in the rotational
period, but with a rate of about 2~s per 100 years.
The authors pointed out that there was an evidence for a second
period of about 5 years attributed to the precession of the axis of rotation.

\subsection[] {A possible long-term variations of 56~Ari}

Three series of 56~Ari observations obtained with SWP~camera cover about
9 years. This fact allows us to investigate a possible systematic
differences between three sets of {\iue} data in order to confirm or not
the second 5 years period found by Adelman et al.~(2001).
Hence, we expect the maxima differences between three sets of
{\iue} data where the amplitudes of flux variations reaches the maxima
values (see Sect.~4.2). One can see that there are not significant
systematic differences between three sets of {\iue} data,
as illustrated by Fig~1.
Generally, taking into account the observational uncertainties of
the fluxes in the spectral region from 1150~\AA~ to 1980~\AA,
it is impossible to confirm the second 5 years period found by
Adelman et al.~(2001).

\begin{figure}
\leavevmode
\begin{center}
\centerline{\includegraphics[width=80mm]{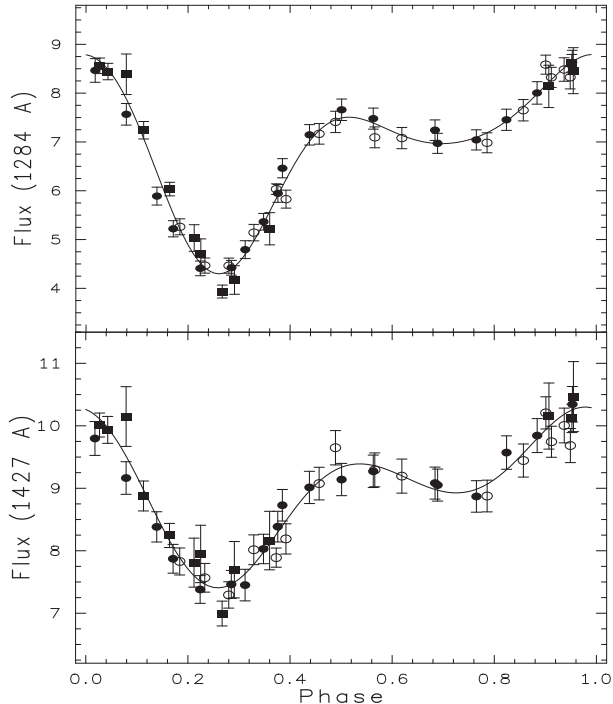}}
\caption{Phase diagrams of the monochromatic light curves in {\units}
 for 56~Ari at $\lambda$1284~\AA~ and $\lambda$1427~\AA. Observations obtained
 during first series (full circles), second one (open circles) and third
 one (full squares). The solid lines are the least square fits.}
\end{center}
\end{figure}

\subsection[] {The accepted period variations of 56~Ari}

The period variation for 56~Ari are quite challengeable.
Nevertheless, in our investigation the phases of the monochromatic light
curves in the far-UV were computed by using ephemerides with constant and
linearily changing period obtained by Adelman et al.~(2001). Additionally,
the ephemeris with constant period obtained by Adelman \& Fried~(1993)
was used as well.
It should be noted that the difference in the ephemeris with constant period
is only in the epoch of the photometric light minimum.
In the most cases the monochromatic light curves
in the far-UV exhibit the maximum of the flux at phase 0.0 with the ephemeris
of Adelman \& Fried~(1993):
\begin{equation}
{\mathrm JD(U,B~min)}=2439797.586+0.727902 E.
\end{equation}

\begin{figure*}
\begin{center}
\centerline{\includegraphics[width=150mm, angle=0]{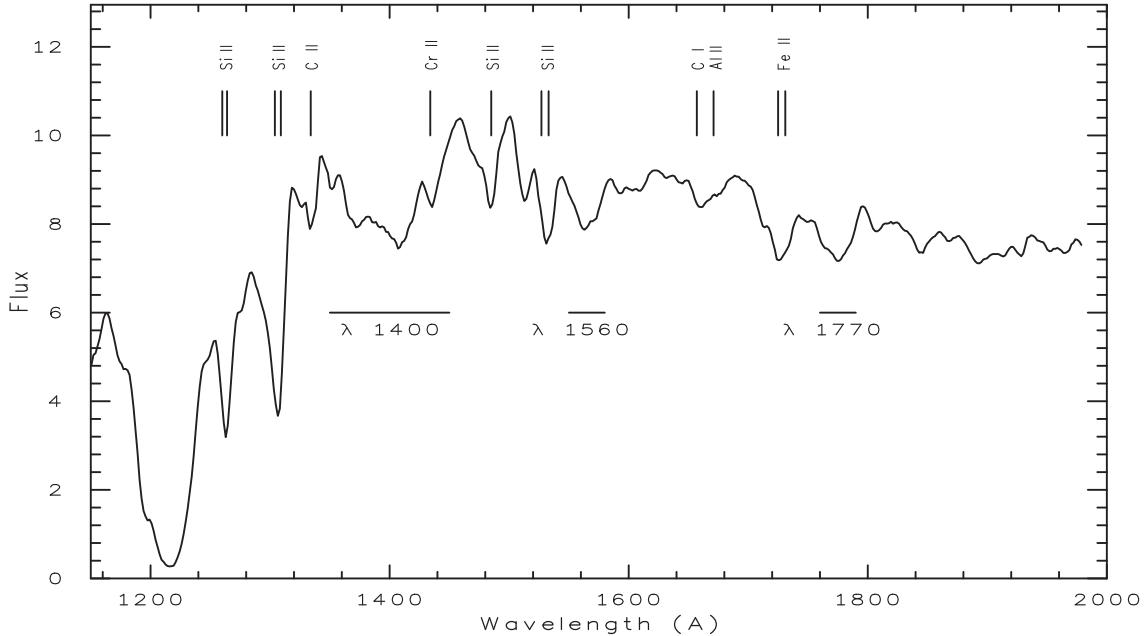}}
\caption{The average energy distribution in {\units} for 56~Ari.
 The prominent spectral lines and features are shown by vertical and
 horizontal lines, respectively.}
\end{center}
\end{figure*}

With this ephemeris, the monochromatic light curves of 56~Ari
show, as a rule, two maxima and two minima in the far-UV spectral region
(Fig.~1). The phases for the observational data were computed
using Eq.~(1).

\section {Data analysis}

To analyse the {\iue} spectra of 56~Ari we used
the linearized least-squares method.
An attempt was made to describe the light curves in a
quantitative way by adjusting a Fourier series.
This method is described by North~(1987) and assumes that
the shape of the curve has the form:
\begin{equation}
F(\lambda, T^{'})=A_{0}(\lambda) + \sum_{i=1}^{n}
A_{i}(\lambda)\cos(\omega~i~T^{'} +\phi_{i}(\lambda)) \label{eq:euler}
\end{equation}
where $T^{'}$=~$T$~-~$T_{0}$ and $\omega$~=~2$\pi$/$P$.
The $T_{0}$ and $P$ are zero epoch and rotational period of
the ephemeris, respectively.
The coefficients $A_{0}(\lambda)$ of the fitted curves
define the average energy distribution over the cycle of the variability.
From several scans distributed over the period one can
produce light curves at different wavelengths.
In all wavelength the data could be
fitted by Fourier series limited to $n$=3, i.e. by the fundamental
frequency and its first two harmonics.
The least-squares fit was applied to separate short-wavelength {\iue}
monochromatic light curves.

\subsection {Identification of the observed lines and features in
the spectrum of 56~Ari}

Figure~2 displays the average energy distribution of 56~Ari over the cycle
of the variability in the spectral region from 1150~\AA\ to 1980~\AA.
Prominent lines are indicated in Fig.~2 with their identification.
One can see in Fig.~2 that Si~II appears as the main absorber with the strong
resonance lines at $\lambda\lambda$1260-64~\AA\AA, 1304-09~\AA\AA, 1485~\AA,
1526-33~\AA\AA, while the weaker one at $\lambda$1808-16~\AA\AA~ is not
detectable at this resolution. According to Artru \& Lanz~(1987),
the strong lines in the spectrum of CP~stars appear
from C~II at $\lambda$1334~\AA\ and Al~II at
$\lambda$1671~\AA, which is close to a strong C~I at $\lambda$1657~\AA\
line. Fe~II form a blend at
$\lambda$1725-31~\AA\AA~ and Cr~II produces line at $\lambda$1434~\AA.

Three large features at $\lambda\lambda$1400~\AA, 1560~\AA\ and 1770~\AA,
which are strongly enhanced in the spectrum of CP stars,
are well seen in spectrum of 56~Ari. Lanz et al.~(1996) gave strong
arguments supporting the idea that the intense autoionization resonanses
of Si~ II could explain
the features at $\lambda\lambda$1400~\AA\ and 1560~\AA\ in the spectrum of
CP stars.
On the other hand, they were unable to identify the depression at
$\lambda$1770~\AA. Another element may cause this strong depression.
It should be noted that the blend at $\lambda$1304-09~\AA\AA~
has two major contributors: the resonance doublet and the autoionising
multiplets (Artru \& Lanz~1987).

\subsection {The monochromatic light variations in pseudo-continuum}

First, it is necessary to fix the continuum in the low dispersion
{\iue} data. This is very difficult in the far-UV due to the lines crowding.
Nevertheless, One can find some high flux points located at the same
wavelengths in several spectra of 56~Ari. It should be noted that such
choice of the high continuum might be a "pseudo-continuum".
But, there is no chance to reach the true continuum, if it occurs at
higher wavelengths.
Several monochromatic light curves in the "pseudo-continuum" at different
wavelengths were formed. The examples of light curves together with the fitted
three-frequency cosine curves are shown in Fig.~3.
Note the vertical scales differ for each part of the figure.
In order to exclude overlaping, the curves at $\lambda\lambda$1502, 1621
and 1906~\AA\AA\ were shifted down to the values of $-2.0$, $-3.0$ and
$-0.5\times${\units}, respectively.

\begin{figure}
\leavevmode
\begin{center}
\centerline{\includegraphics[width=100mm]{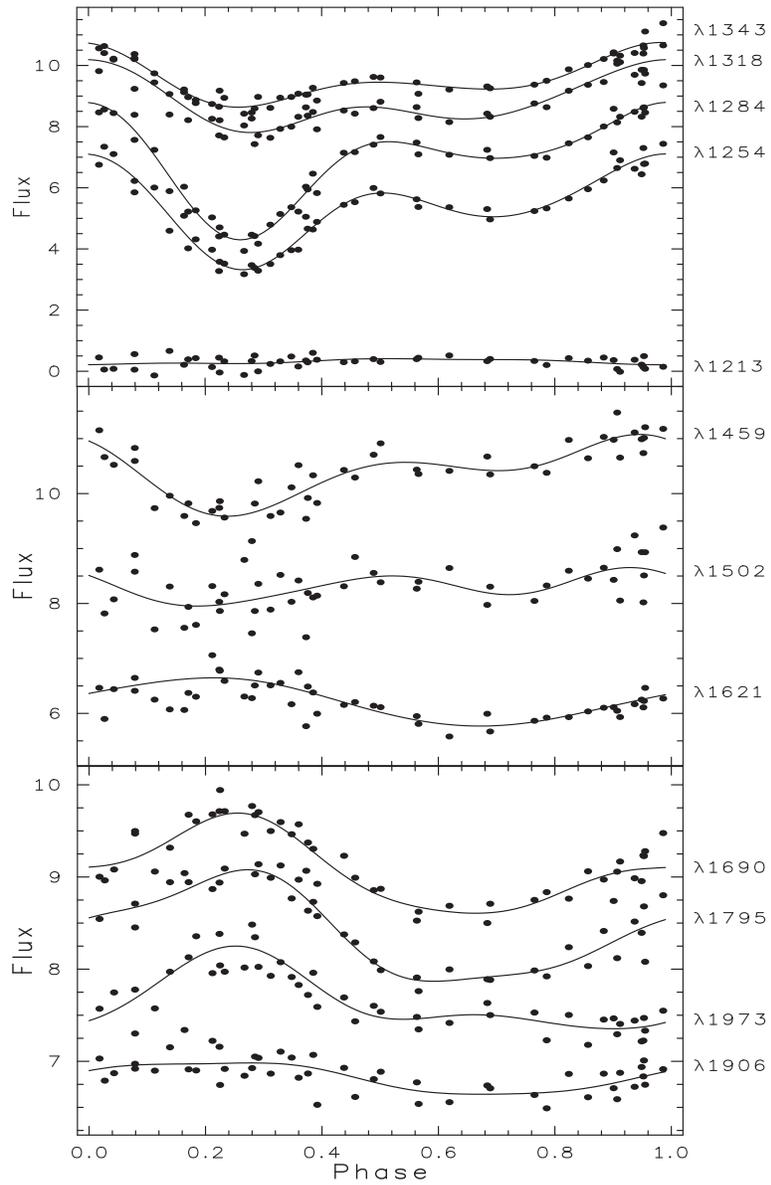}}
\caption{Phase diagrams of the monochromatic light curves in {\units}
  for 56~Ari. To exclude the overlap the vertical shift on the constant value
  of some curves was used (see text). The solid lines are the least square fits.}
\end{center}
\end{figure}

The light curves of 56~Ari change their shape with wavelength.
All curves in the spectral region with $\lambda~<$~1600~\AA\
have a similar shape: the primary maximum at phase 0.0, a deep minimum
at phase 0.25, while a secondary maximum and minimum around the phases 0.5
and 0.65, respectively. It should be noted, that Stepie{\' n} \&
Czechowski~(1993) obtained a deep minimum at phase 0.8-0.9 and another one
at phase 0.3-0.4 (see their Fig.~2). This disagreement may arise
because they computed the phases with the old ephemeris given by
Hardie \& Schroeder~(1963).
The double-wave light variations in this spectral region are
in antiphase to the variations in the visual spectral region
(Adelman \& Fried 1993).
The exception is the core of $L_{\alpha}$ line at $\lambda$1213~\AA\
where the amplitude of light variations is practically zero over the period
of rotation. Another rapidly rotating CP~star CU~Vir
displays the same behavior for the core of $L_{\alpha}$ line (Sokolov~2000).

The amplitude of the minimum at phase 0.25 decreases with increasing
wavelength, but at $\lambda$~$\sim$~1600~\AA\ and beyond this minimum
is replaced by a maximum. This maximum is seen up to $\lambda$1980~\AA,
except for the cores of the blends at $\lambda\lambda$1671, 1727~\AA,
the feature at $\lambda$1770~\AA\ and aroud $\lambda$1900~\AA, where
the maximum disappears again (see Fig.~4).
The second minimum decreases with increasing wavelength and then
disappears at $\lambda$~$\sim$~1973~\AA.
At $\lambda$1621~\AA\ the amplitude of the two features is the same and,
as a result, the cosine wave is seen at this wavelength.
A very complex monochromatic light variations are detected in
the far-UV of 56~Ari. The spectrophotometric behavior in
the far-UV of the star CU~Vir is quite similar.
In the spectrum of CU~Vir the first minimum is replaced by the maximum at
$\lambda$1611~\AA\ and the second minimum disappears at
$\lambda$1962~\AA\ (Sokolov~2000).

\begin{figure*}
\leavevmode
\begin{center}
\centerline{\includegraphics[width=150mm, angle=0]{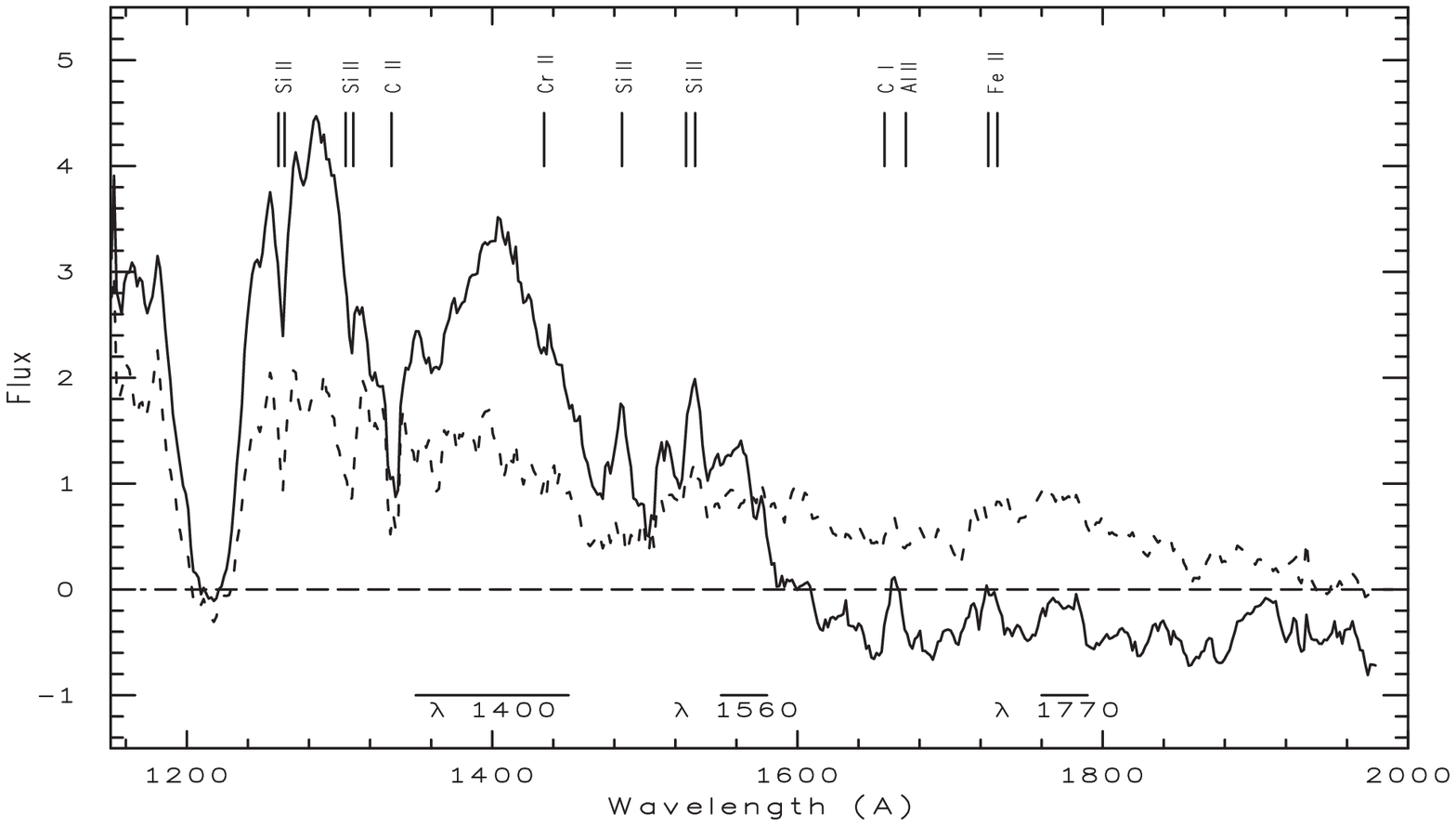}}
\caption{The differences of energy distributions of 56~Ari
 in {\units} for three phases.
 The solid line shows the difference between phases 0.0 and 0.25;
 The dashed line shows the differences between phases 0.0 and 0.65.
 Positive values of these differences mean that the star is brighter
 at phase 0.0.}
\end{center}
\end{figure*}

\subsection{The null wavelength regions}

The spectrum of 56~Ari is highly variable in the wavelength interval
from 1150~\AA\ to 1980~\AA\ over the whole rotational period.
This indicates that even when monochromatic light curves beyond a given
wavelength are in antiphase to those in shorter wavelengths, we do
not observe a truly "null wavelength region" where the monochromatic
light curve has a zero amplitude.
Instead, it is possible to establish the wavelengths where the fluxes
of the primary and secondary minima transform into the maxima.
We will call this a local "null wavelength regions".
Thus, the differences of the energy distributions between phases 0.0
and two minima reflect a dependence of the amplitude of light variations
at phases of minima on wavelength.

Using the coefficients of the fitted curves in each wavelength,
the energy distributions at phases 0.0, 0.25 and 0.65 were computed.
Figure~4 presents the differences in the fluxes between phases 0.0 and 0.25
and between phases 0.0 and 0.65 of the rotational period of 56~Ari.
It is apparent from Fig.~4 that the first minimum at phase 0.25 is seen up to
$\lambda$1586~\AA. The local "null wavelength region" for the first minimum
is located in the spectral region between $\lambda$1586~\AA\ and
$\lambda$1608~\AA, where the fluxes at phases 0.0 and 0.25 are equal.
For $\lambda~>$~1608~\AA\ the first minimum is replaced by the maximum,
except for the cores of Al~II at $\lambda\lambda$1671, Fe~II at
1727~\AA\AA, the feature at $\lambda$1770~\AA\ and
aroud $\lambda$1900~\AA, where the local "null wavelength region"
for the first minimum is seen again.
On the other hand, the differences of the fluxes at
phases 0.0 and 0.65 slowly decrease with increasing wavelength.
The second minimum at phase 0.65 is seen up to the end
of SWP spectrum, as illustrated by Fig.~4.
The local "null wavelength region" for the second minimum is reached
in the spectral region between $\lambda$1938~\AA\ and $\lambda$1980~\AA,
where the fluxes at phases 0.0 and 0.65 are equal.
Note that the amplitudes of light variations at phases 0.25 and 0.65 reach
the minimum values at $\lambda\sim$1340~\AA\ and in the cores of Si~II
resonance lines at $\lambda$1260-64~\AA\AA\ and $\lambda$1304-09~\AA\AA.

For silicon-rich stars with the effective temperature of about 10000 - 15000~K,
Si~II becomes the major source of continuum
opacity in spectral region with $\lambda~<$~1600~\AA.
Lanz et al.~(1996) have shown that the Si~II opacity is
comparable to the H~I opacity at many frequencies and can produce
most UV features of these stars.
The variability of the Si~II continuum opacity could explain the deep
minimum of the flux at phase 0.25 and its redistribution in longer
wavelengths for the star 56~Ari.
The amplitude of the first minimum at phase
0.25 reaches the maxima values of 45\% and 65\% in the large feature at
$\lambda$1400~\AA\ and in the continuum at $\lambda$1280~\AA, respectively.
The theoretical continuum spectrum which includes the Si~II continuum
opacity shows at this wavelengths the maxima values of the total
continuum opacity (see Fig.~5 from Lanz et al.~(1996)).
Also, the minimum of amplitude (11.9\%) at $\lambda$$\sim$1340~\AA\
corresponds to the wavelength, where Si~II continuum opacity is minimal.
On the other hand, the differences of the fluxes at
phases 0.0 and 0.65 slowly decrease with increasing wavelength.
Probably, there are other blanketing source in the far-UV of 56~Ari.
First of all, we should expect a close correlation between light variations
and a non-uniform distribution of silicon on the surface of 56~Ari.

\subsection {Variation of the broad features}

Two broad features at $\lambda\lambda$1400, 1560~\AA\AA~ are well seen in
spectrum of 56~Ari, as illustrated by Fig.~2.
Remembering the identifications detailed in Sect.~4.1, we have also
included in our investigation the unidentified diffuse depression
around $\lambda$1770~\AA\ (see Fig.~2).

To measure the total absorption in the broad features at
$\lambda\lambda$1400, 1560 and 1770~\AA\AA~ we have formed the photometric
indices $a_{1400}$, $a_{1560}$ and $a_{1770}$, expressed in magnitudes:
\begin{eqnarray}
a_{1400}={1 \over 2}(m_{1342}+m_{1441})-m_{1397},  \nonumber \\
a_{1560}={1 \over 2}(m_{1544}+m_{1584})-m_{1562},  \label{eq:euler} \\
a_{1770}={1 \over 2}(m_{1752}+m_{1795})-m_{1770},  \nonumber
\end{eqnarray}
where all filters are 10~\AA\ wide for the $a_{1400}$ index and all filters
are 7~\AA\ wide for the $a_{1560}$ and the $a_{1770}$ indices.
These indices are analogous to the $a_{1400}$ index of Shore \& Brown~(1987).
The equivalent width measurement is not the best way of determining
the strength  of the broad feature. Shore \& Brown~(1987) noted that there is
more uncertainty associated with the equivalent width measurements than
the impersonal photometry due largely to judgement exercised  in
the placement of the continuum.
Therefore, that the equivalent widths are correlated with $a_{1400}$ index
is a useful result, since it means that one can rely on the automated
procedure for measuring the total absorption at 1400~\AA~ (see Fig.~4 from
Shore \& Brown~(1987)).

\begin{figure}
\leavevmode
\begin{center}
\centerline{\includegraphics[width=100mm, angle=0]{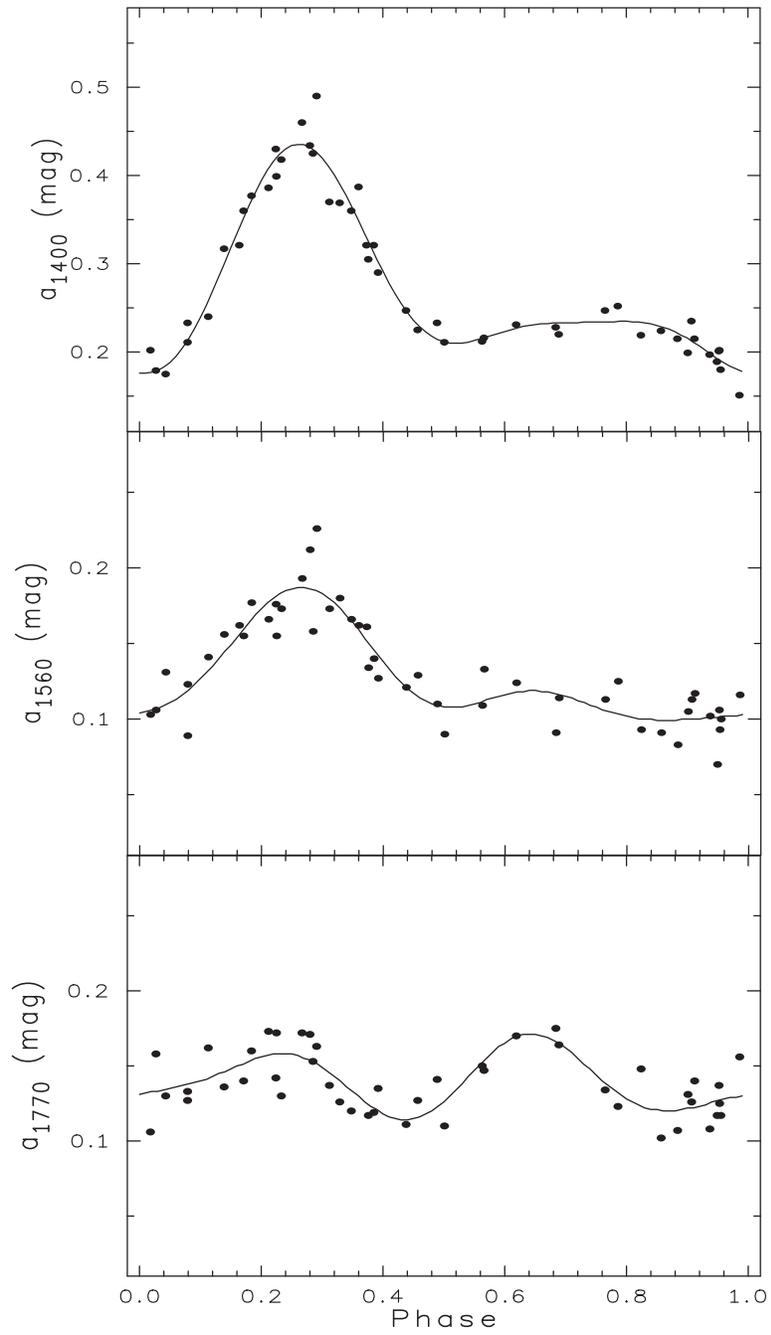}}
\caption{Phase diagrams of the broad features in the far-UV spectral
 region of 56~Ari. Note different vertical scales for each part of
 the figure. The solid lines are the least square fits.}
\end{center}
\end{figure}

\begin{figure}
\leavevmode
\begin{center}
\centerline{\includegraphics[width=100mm, angle=0]{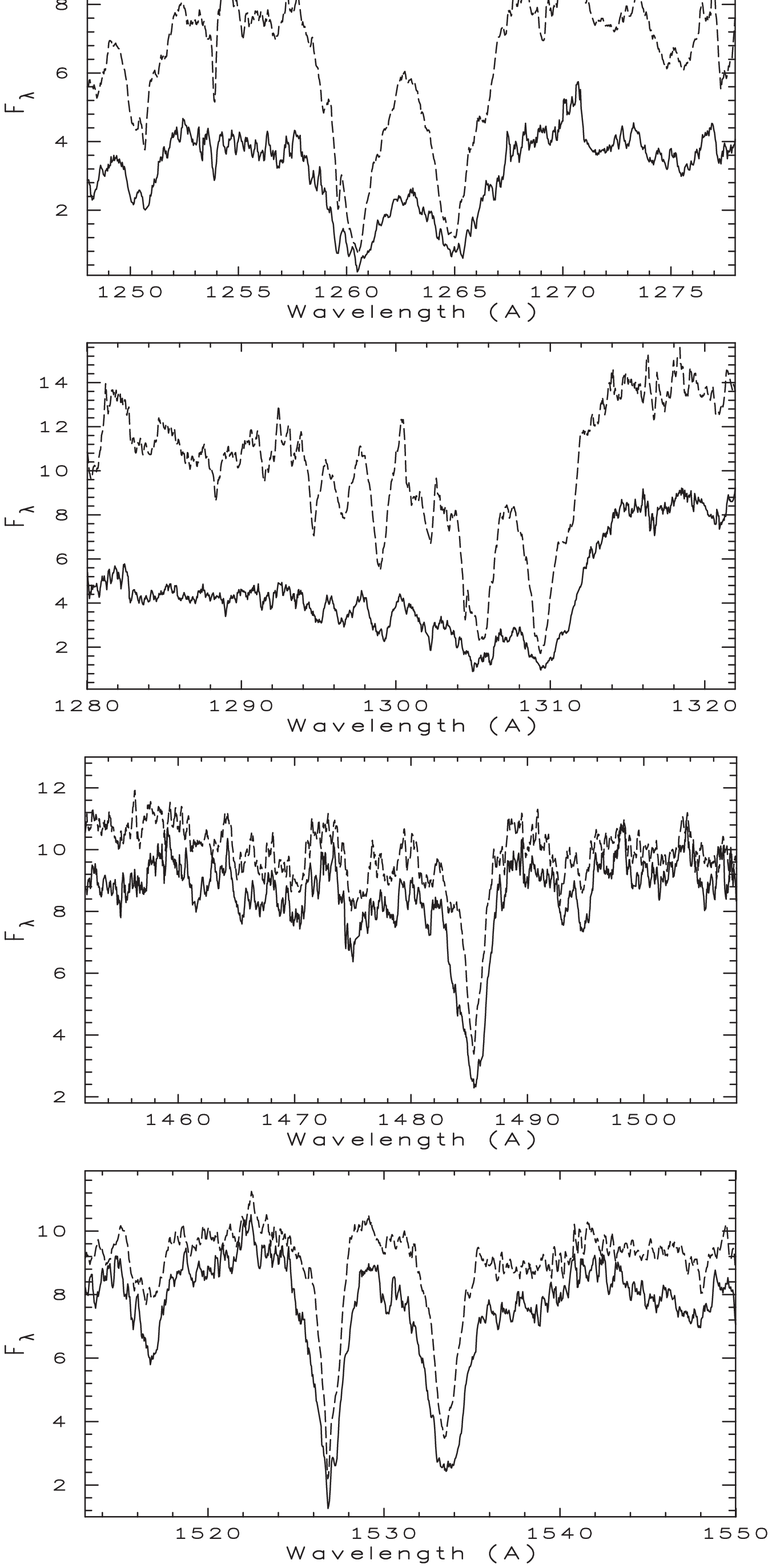}}
\caption{The high-dispersion spectra of 56~Ari in the regions of
 Si~II resonance lines at the phases 0.0 (dash line) and 0.25 (full line).}
\end{center}
\end{figure}

Figure~5 shows the variations of the measured total absorption in the broad
features at $\lambda\lambda$1400, 1560 and 1770~\AA\AA~ versus
the rotational phase. The solid lines represent least-squares fits by
three-frequency cosine functions.
As one can see from Fig.~5, the deviation of the observational data
from the fitted curve may be significant.
The maxima deviations of the observational data from the fitted curves
of the photometric indices $a_{1400}$, $a_{1560}$
and $a_{1770}$ are equal to 0.064~mag, 0.041~mag and 0.028~mag, respectively.
The mean standard deviation of the residuals scatter around the fitted
curve ($\sigma_{res}$) were computed.
For fitted curves of the photometric indices $a_{1400}$, $a_{1560}$
and $a_{1770}$ the $\sigma_{res}$ are equal to 0.019~mag, 0.016~mag
and 0.015~mag, respectively.

We should expect a close correlation
between light variations and the variations of the UV features.
In fact, the variability of the broad features is rather complex.
Thus, the $a_{1400}$ index shows the maximum around the 0.25 phase
with amplitude of $\sim$0.25~mag. Note that, within errors, this index does
not vary at phases 0.5-1.0. The $a_{1560}$ index also shows the maximum
around the 0.25 phase but with amplitude of $\sim$0.1~mag and it does
not vary at phases 0.5-1.0 either. On the other hand, the $a_{1770}$ index
shows the primary maximum at phase 0.25 and the secondary maximum
at phase 0.65. Both maxima have very small ($\sim$0.05~mag) amplitude.
Quantitative comparison  of the monochromatic light curves
with variations of the broad features shows that only variations of
the primary minimum of light curves are in antiphase to the variations
of the Si~II features as if Si~II were concentrated in one spot only.
However, the monochromatic light curves are completely in
antiphase to the $a_{1770}$ index variations.

\section {Discussion}

Additional sources of variable opacity must be
involved in order to explain the existence of the second minimum at
phase 0.65 of the monochromatic light curves.
At the first time, Panek~(1982) noted that flux redistribution by no
single element can explain the light variations at all wavelengths,
although Si~II dominates in the range $\lambda\lambda$1150-1500~\AA.
Stepie{\' n} \& Czechowski~(1993) investigated the influence
of silicon, helium, and iron on the monochromatic light variations in
the spectrum of 56~Ari. The authors noted that more quantitative comparison
of the light variations with the variations of Si~II $\lambda$4128~\AA\
shows some inconsistency. Also, they concluded that the variations of helium
seem to have very little influence on the light variations and
the assumed strict in phase variation of iron with silicon, as observations
suggest, is probably an over-interpretation of the existing data.

Two different causes contribute to flux variability in the spectra of
CP stars: variable line blocking and changes in the continuum flux.
To illustrate the importance of the second cause in the spectrum of 56~Ari,
we compare the energy distribution at the phases 0.0 and 0.25 using
high-dispersion spectra. In this way, the regions of the Si~II resonance
lines were selected. The mean high-dispersion spectra were computed
at the phases 0.0 and 0.25 using {\iue} images SWP~39489, SWP~39678,
SWP~39680 and SWP~22331, SWP~22333, SWP~39685, respectively.
Figure~6 exhibits the mean high-dispersion spectra of 56~Ari in
the regions of Si~II resonance lines at the phases 0.0 and 0.25.
The comparison of these two energy distributions show that in the regions
of Si~II resonance lines at $\lambda$1260-64~\AA~ and $\lambda$1304-09~\AA~
the variability of the Si~II continuum opacity could explain the deep
minimum  of the flux at phase 0.25 for the star 56~Ari.
Note that the exception is the cores of these Si~II resonance lines
where the variability of the flux is negligible.
On the other hand, in the regions of Si~II resonance lines at
$\lambda$1485~\AA~ and $\lambda$1526-33~\AA~ the variability of the flux
in the continuum is significantly less, but the profiles of these Si~II
resonance lines are more variable.
This result supports the idea to use the low-dispersion spectra of 56~Ari
in order to investigate the variability of the monochromatic light curves.

\section {Conclusions}

The archival {\iue} spectrophotometric observations of 56~Ari permit
to analyse the light variations in the far-UV spectral region from 1150~\AA\
to 1980~\AA\ and to compare them with the variations in the cores of
the spectral features and lines.
Although the double-wave light variations in this spectral region are in
antiphase to those in the visual region,
the spectrum of 56~Ari is highly variable in the wavelength interval
from 1150~\AA\ to 1980~\AA\ over the whole rotational period.
The first minimum of the light curves at phase 0.25 is replaced by
the maximum for $\lambda$~$>$~1608~\AA, but the second minimum of
the light curves at phase 0.65 is seen up to the end of SWP spectrum.
As a result, the monochromatic light curves continuously change their
shape with wavelength.
The amplitudes of light variations reach the minimum
values in the core of $\mathrm{L}_{\alpha}$ line,
where the flux is formed in the outer layers of the stellar atmosphere.

The variable broad features in the far-UV connected with a
non-uniform distribution of silicon over the surface of 56~Ari
influence substantially the light variations in the UV.
The anticorrelation between the light variations in the far-UV
and the silicon features intensity variations is caused by extra
blocking of the flux in the far-UV and its redistribution  in
the longer wavelengths.
Moreover the comparison of these two mean high-dispersion spectra of 56~Ari
show that in the regions of Si~II resonance lines at $\lambda$1260-64~\AA~
and $\lambda$1304-09~\AA~ the variability of the Si~II continuum opacity
could explain the deep minimum  of the flux at phase 0.25 for this star.

\end{document}